\documentclass[preprint2]{emulateapj}
 
\def\lesssim{\buildrel < \over {_{\sim}}}
\def\gtrsim{\buildrel > \over {_{\sim}}}

\newcommand{\dmrho}{\mathrm{GeV cm^{-3}}} 
 
\shorttitle{Evolution of the first stars with DM burning}
\shortauthors{Yoon, Iocco \& Akiyama}
 
\begin{document} 
 
 
\title{Evolution of the First Stars with Dark Matter Burning}

 
\author{Sung-Chul Yoon \altaffilmark{1}} 
\affil{UCO/Lick Observatory, Department of Astronomy and Astrophysics, University of 
  California, Santa Cruz, CA 95064, USA, \email{scyoon@ucolick.org}}
 
\author{Fabio Iocco \altaffilmark{2}} 
\affil{INAF -- Osservatorio Astrofisico di Arcetri, Largo Enrico Fermi 
5, Firenze 50125, Italy, \email{iocco@arcetri.astro.it}}
 
\and 
 
\author{Shizuka Akiyama \altaffilmark{3}} 
\affil{Stanford Linear Accelerator Center and 
Kavli Institute for Particle Astrophysics and Cosmology, 
Stanford, CA 94305, USA, \email{shizuka@slac.stanford.edu}} 
 
 
 
 
\begin{abstract} 
Recent theoretical studies have revealed the possibly important role of the  capture and
annihilation process of weakly interacting massive particles (WIMPs) for the first stars.  Using new
evolutionary models of metal-free massive stars, we investigate the
impact of such ``dark matter burning'' for the first stars in
different environments of dark matter (DM) halos, in terms of the ambient WIMP density
($\rho_\chi$).  We find that, in agreement with existing literature, stellar life times can be
significantly prolonged for a certain range of $\rho_\chi$ (i.e., $10^{10} \la \rho_\chi [\dmrho]
\la 10^{11}$ with the current upper limit for the spin-dependent elastic scattering cross section
$\sigma_0^\mathrm{SD} = 5 \times 10^{-39} \mathrm{cm^2}$).  This greatly
enhances the role of rotationally induced chemical mixing in rotating stars, in favor of abundant
production of primary nitrogen, massive helium stars and long gamma-ray bursts, from the first
stars.  We also find that stars with $\rho_\chi > 2 \times 10^{11}~\dmrho$ may not
undergo nuclear burning stages, confirming the previous work, and that ionizing photon fluxes from
such DM supported stars are very weak.  Delayed metal enrichment and slow reionization in the early
universe would have resulted if most of the first stars had been born in DM halos with  such high
$\rho_\chi$, unless it had been lowered significantly below the threshold for efficient DM burning
on a short time scale.  
\end{abstract} 
 
\section{Introduction} 
 
In the current framework of the $\Lambda$CDM cosmological model, the possibility that dark matter
(DM) annihilation effects may impact stellar evolution has recently received renewed attention.
After the  original works of Bouquet, Dearborn, Freese, Gould, Griest, Krauss, Olive,Press, Raffelt,
Renzini, Salati, Silk, Spergel, Srednicki and Wilczek in the`80s and early `90s, several authors
have recently re--examined the effects that DM, if made of weakly interacting massive particles
(WIMPs), would have on compact objects  \citep{Moskalenko:2007ak, Bertone:2007ae} and on the
zero-age main sequence of low--mass stars \citep{Fairbairn:2007bn}.  This exciting activity has been
motivated by  a double scope: any peculiar and distinguishable feature of WIMP  annihilation on
observable stellar quantities is extremely precious in the ``quest'' for dark matter evidence; on
the other hand, all possible  effects impacting the life of celestial objects must be taken into
account by astrophysicists in the current precision era.  
 
In particular, the first stellar episode at high redshift occurs under very  different
conditions from those in the present universe.  The higher concentration of dark matter,  the short
Hubble time which prevents DM self--annihilation  from severely affecting the central density, and
the characteristic formation of a single PopIII star in the center of  the halo are the most
favorable conditions for DM annihilation effects to be very efficient in the first stars.  In
their pioneering work,  \citet{Spolyar:2008qv} first found the possibility of very high DM
density in primordial halos at high redshift to such an extent that energy released from DM
annihilation at the center may halt the gravitational collapse of the baryonic cloud, calling such a
DM powered object a {\it dark} star.  \citet[hereafter I08]{Iocco:2008xb} and \citet{Freese:2008ur}
also noticed that WIMP capture is most efficient in Population III stars.  More recently,
\citet{freese08a, freese08b} and \citet{Iocco:2008rb} further investigated the role of annihilation
of adiabatically contracted DM in the formation of the first stars.  \citet{Iocco:2008rb} have also
followed the evolution from the pre-main sequence phase to helium exhaustion in the presence of WIMP
capture and annihilation (hereafter, DM burning), showing that this can severely delay the evolution
of pre--MS objects in the early universe, as well as extend their MS lifetimes. 

All of these studies motivate us to explore possible consequences of DM burning for the final fate
of the first stars and their feedback effects on the evolution of the early universe, even if the
role of DM annihilation in the formation of the first stars still remains subject to many uncertainties
\citep[see][for a recent review]{FS3proc}.  In this Letter, we address the issue by discussing the
evolution of the first stars of $20\le M/\mathrm{M_\odot} \le 300$ up to the carbon burning stage
(Sect.~\ref{sect:results}).  We also investigate the interplay of rotation with DM burning in the
evolution of the first stars of $100~\mathrm{M_\odot}$, given the particular importance of rotation
for the evolution and deaths of metal poor massive stars \citep[e.g,][]{Meynet08, Yoon08}.
Implications of our results for the history of reionization in the early universe are briefly
discussed (Sect.~\ref{sect:discussion}). 

\section{Physical Assumptions and Results}\label{sect:results} 
 
We have implemented the DM capture and annihilation process in a hydrodynamic  stellar evolution
code, following \citet{gould87}.  The DM capture rate $C_*$  is calculated using Gould's equations
as reported in Eqs. (1,2) of I08.  Throughout this Letter we assume the DM--baryon scattering cross
section $\sigma_0$  is $5\times10^{-39}~\mathrm{cm^2}$ for the spin-dependent scattering, to which
only hydrogen is sensitive, and $10^{-43}~\mathrm{cm^2}$ for the spin-independent one. These
correspond to the current upper limits of WIMP direct detection
search~\citep{Desai:2004pq,Angle:2008we}.  We adopt the same values for other parameters as those
used in I08 to calculate $C_*$, except for the ambient DM density (see below).  The WIMPs captured
by a star eventually reach thermal equilibrium  with the gas, in a configuration dictated by the
gravitational potential: the consequent DM density can be given by $n_\chi(r) = n_\chi^\mathrm{c}
\exp(-r^2/r_\chi^2)$, where $r_\chi = c ( 3kT_\mathrm{c}/2\pi
\rho_\mathrm{c}m_\mathrm{\chi})^{1/2}$~\citep{Griest87}.  Here $T_{c}$ and $\rho_{c}$ are
temperature and density at the stellar center, $c$ and $k$ are the speed of light and Boltmann's
constant, respectively.  The energy generation rate due to DM annihilation is given by
\begin{equation} \epsilon_\chi (r) = \frac{2}{3} <\sigma v>n_\chi^2(r) m_\chi
~~[\mathrm{erg~cm^{-3}~s^{-1}}]~~,  \end{equation} where $m_{\chi}$ is the mass of the DM particle;
we assume it to be m$_\chi$=100 GeV, which is often taken as a fiducial value in astrophysical
studies for DM search.  We use $<\sigma v> = 3 \times 10^{-26}~\mathrm{cm^3~s^{-1}}$, the value best
fitting the relic DM abundance ~\citep[see, e.g., ][for a recent review]{Bertone05}.  The factor
$2/3$ is to consider that part of the energy is carried away by neutrinos.  We  consider the time
dependent evolution of the total number of WIMPs ($N_\mathrm{tot}$) in the star,  and the number of
thermally relaxed WIMPs in the core ($N_\mathrm{th} := \int n(r) dV$) by the following equations in
order to normalize $n(r)$: \begin{eqnarray} \frac{dN_\mathrm{tot}}{dt} & = & C_* -  \int n(r)^2
<\sigma v> dV, ~\mathrm{and} \\ \frac{dN_\mathrm{th}}{dt} & = & \Gamma_\mathrm{th} -  \int n(r)^2
<\sigma v> dV~, \end{eqnarray} where $\Gamma_\mathrm{th}$ is the thermalization rate that can be
approximated by  \begin{equation} \Gamma_\mathrm{th} = \frac{N_\mathrm{tot} -
N_\mathrm{th}}{\tau_\mathrm{th}},~~~\tau_\mathrm{th} =
\frac{4\pi}{3\sqrt{2G}}\frac{m_\chi}{\sigma_0}\frac{R^{7/2}}{M^{3/2}}~.  \end{equation} See
\citet{Iocco:2008rb} for detailed discussion on the thermalization time scale $\tau_\mathrm{th}$.
Our models show that equilibrium between $C_*$ and $\Gamma_\mathrm{th}$ is well maintained up to the
core helium burning phase (see Fig.~\ref{fig2} below). 

We consider several different values for the ambient DM energy density $\rho_\chi$ ranging from 0 to
$2\times10^{12}~\mathrm{GeV~cm^{-3}}$.  For the initial composition of the first stars, we assume
the mass fractions of $^1\mathrm{H}$, $^4\mathrm{He}$ and $^3\mathrm{He}$ to be 0.76, 0.23999, and
0.00001, respectively.  The mass loss rate from metal free stars is assumed to be zero if the
Eddington factor $\Gamma_\mathrm{E}$ is smaller than 0.84 and $10^{-14}~\mathrm{M_\odot~yr^{-1}}$
otherwise, following~\citet{Krticka06}.  If a star becomes a helium star by rotationally induced
mixing (see below),  we assume that the mass loss rate from such a metal free Wolf-Rayet (WR) star
is the same as that from a corresponding WR star at $Z=10^{-6}$,  as implied by \citet{Vink05}.  The
code also implements the transport of chemical species and angular momentum due to rotationally
induced hydrodynamic instabilities and the Spruit-Tayler dynamo.  Other details on the stellar
evolution code are described in \cite{Yoon06}.  

\begin{figure}[t] \epsscale{1.0} 
\plotone{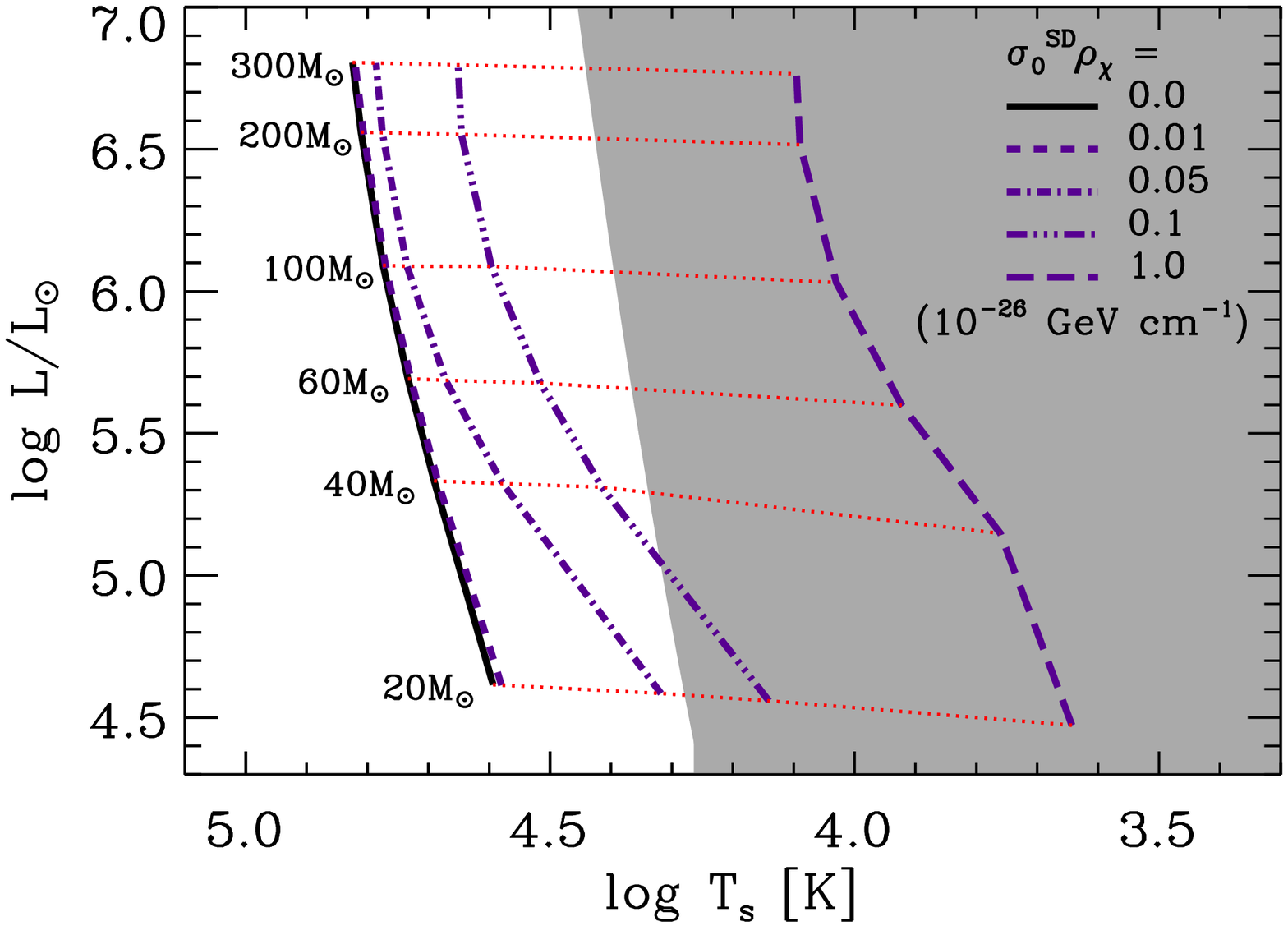} 
\plotone{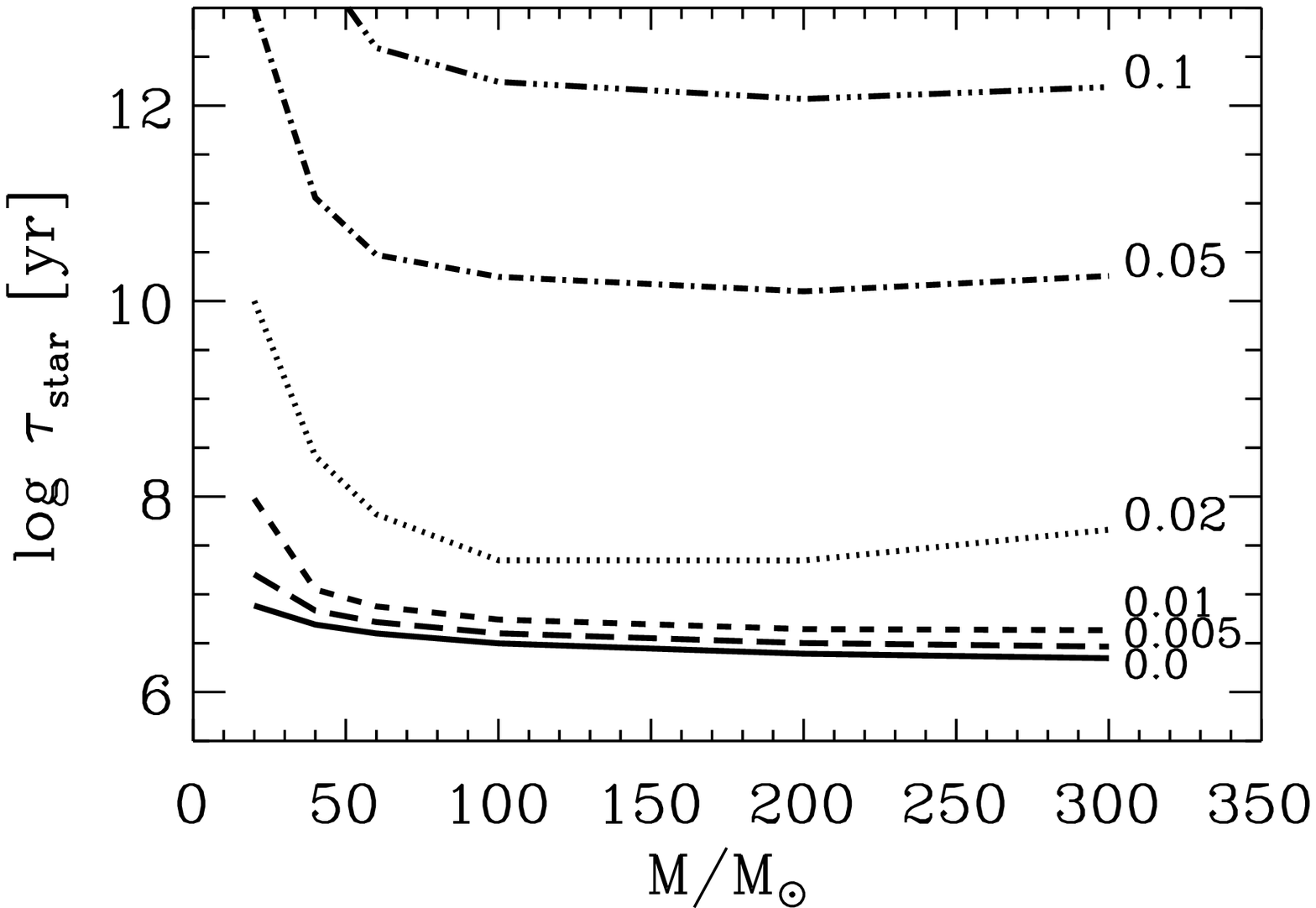} 
\caption{\emph{Upper panel:} HR
diagram of the non-rotating first star models on the ZAMS of different masses  for different
adopted values of $\sigma_0^{SD} \rho_\chi$ as indicated by the labels.  Here $\sigma_0^{SD}$ is the
spin dependent WIMP scattering cross section, and $\rho_\chi$ is the ambient WIMP density.  Stars in
the grey shaded region are supposed to be only powered by DM burning, without nuclear reactions
(i.e., $T_\mathrm{c} < 10^7~\mathrm{K}$). We use $\sigma_0^\mathrm{SD} =
5\times10^{-39}~\mathrm{cm^2}$ \emph{Lower panel:} Life times of the non-rotating first stars as a
function of the initial mass for different values of $\sigma_0^\mathrm{SD} \rho_\chi$ given in the
unit of $10^{-26}~\mathrm{GeV cm^{-1}}$.  These life times are obtained from stellar models up to
carbon burning for $\rho_\chi \le 4 \times 10^{10}~\dmrho$ (i.e.,$\sigma_0^\mathrm{SD} \rho_\chi \le
0.02 \times 10^{-26}~\mathrm{GeV cm^{-1}}$), while only approximate estimates are given for
$\rho_\chi \ge 10^{11} \dmrho$ (i.e.,$\sigma_0^\mathrm{SD} \rho_\chi \ge
0.05\times10^{-26}~\mathrm{GeV cm^{-1}}$) .  
}\label{fig1}  
\end{figure}

Fig.~\ref{fig1} shows the HR diagram of the constructed non--rotating  first star models on the zero
age main sequence (ZAMS),  for different values of $\rho_\chi$.  Note that the luminosity for a
given mass does not significantly  change with varying $\rho_\chi$. This represents the well--known
nature of the mass--luminosity  relation, which is largely independent of any particular mode of
energy generation~\citep{Kippenhahn90}.  The stellar structure is adjusted such that the ratio of
the nuclear to the DM luminosity decreases with increasing $\rho_\chi$, while  the total luminosity
remains nearly constant, leading to prolonged lifetimes of the stars \footnote{For a given
$\rho_\chi$, our models give longer life  times than those by \citet{Iocco:2008rb}.  This is because
these authors used an approximation for  the DM capture rate $C_*$, while we integrate Eq (2) of I08
over the entire stellar structure.} (Fig.~\ref{fig1}).  If $\rho_\chi$ is above a critical value
($\rho_\mathrm{\chi, crit} \approx  2\times10^{11}~\dmrho$, see Fig.~\ref{fig1}) at a given stellar
mass the central density and temperature decrease to such an extent that the stars would live
forever on  the ZAMS, having no nuclear reactions. 
This is in qualitative agreement with
\citet{Iocco:2008rb} who find that  pre--MS evolution is ``frozen'' before reaching the ZAMS at high
enough $\rho_\chi$; and with \citet{Fairbairn:2007bn}, who find (for M$\leq$4$M_\odot$) that stars
move rightward of the ZAMS locus on the HR, and eventually join the Hayashi track,  if ``fed'' with
increasing DM annihilation luminosities.

\begin{figure}[t] 
\epsscale{1.00} 
\plotone{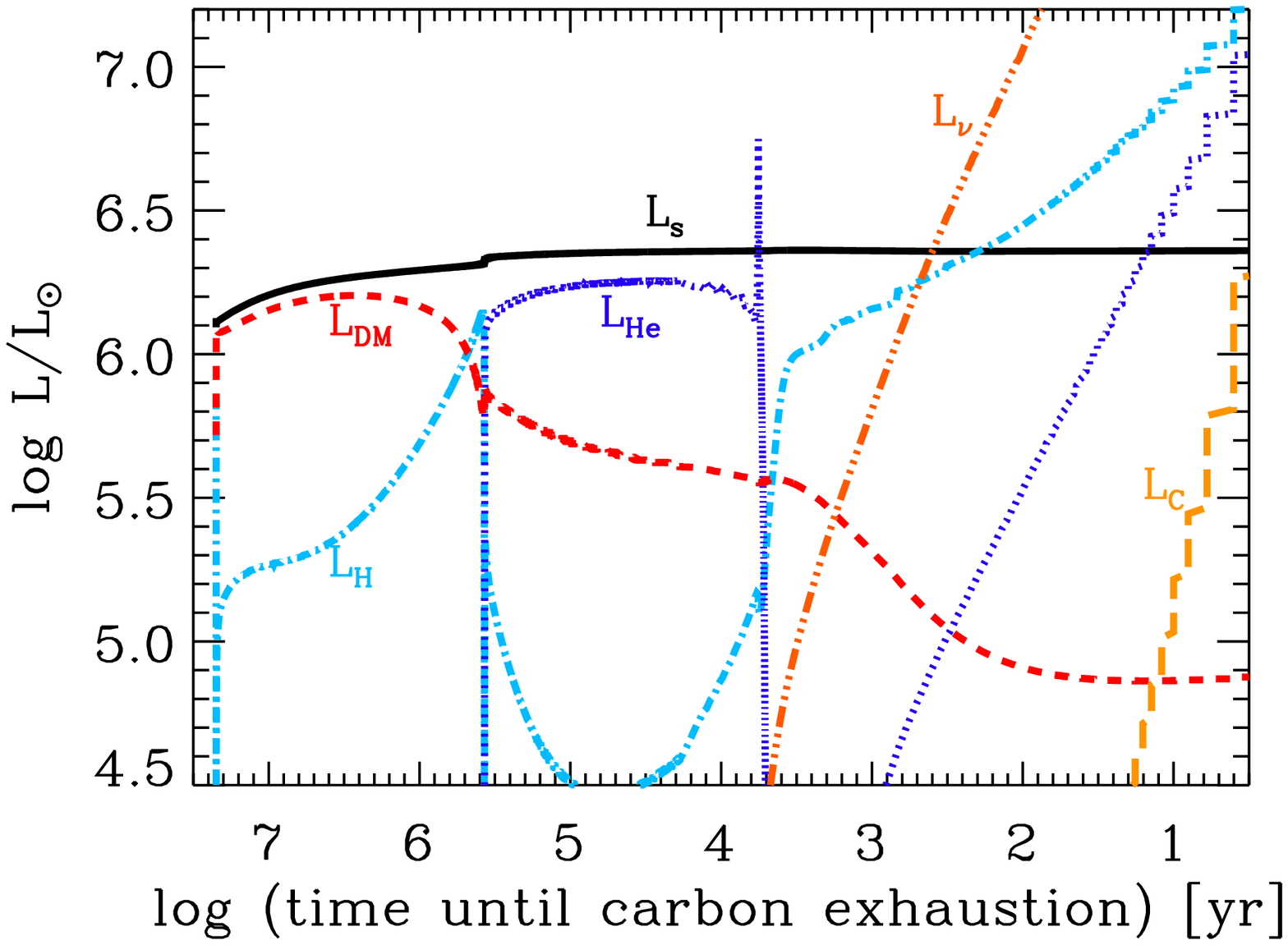} 
\plotone{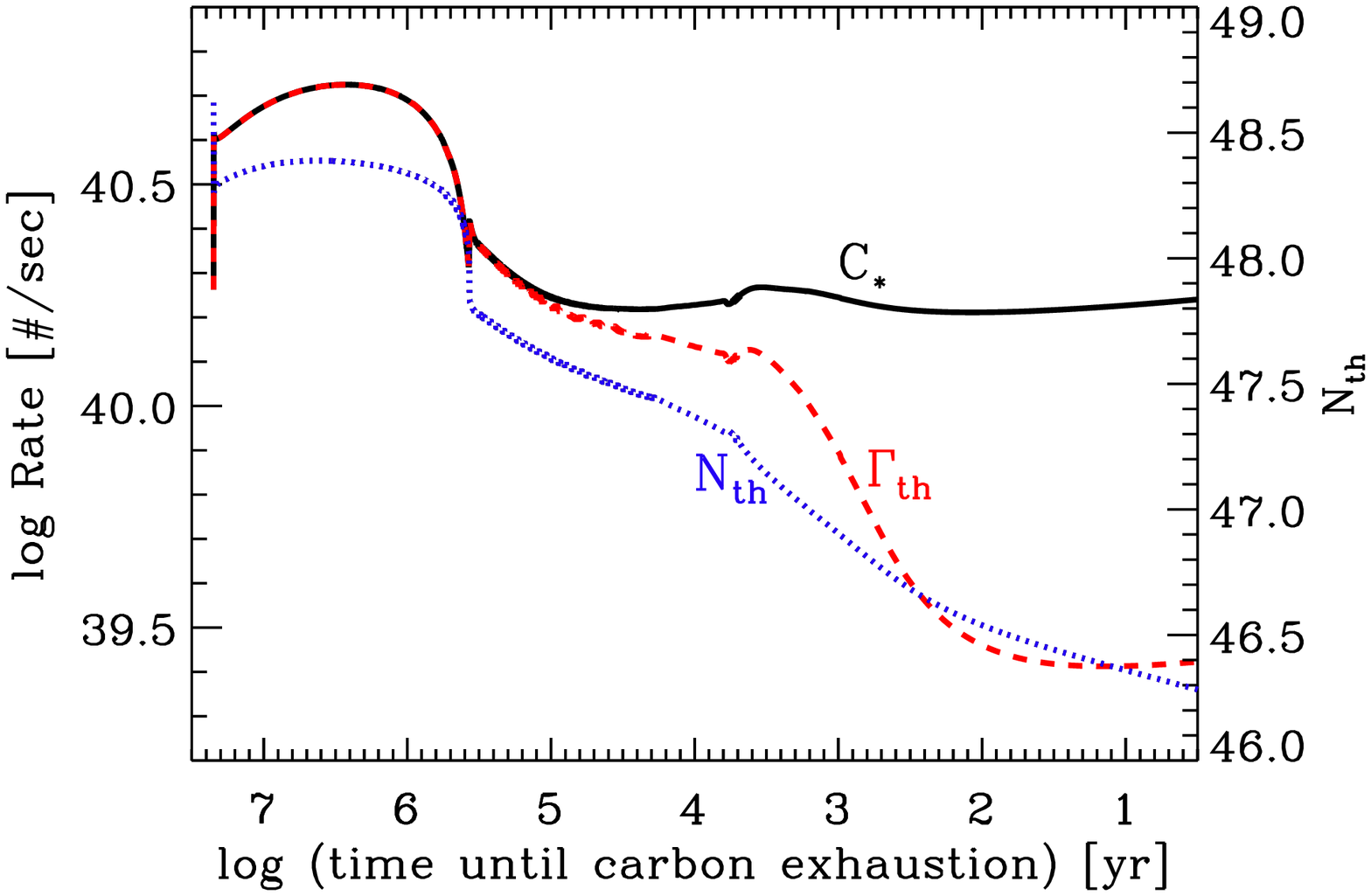} 
\caption{
\emph{Upper Panel:} Evolution of the surface luminosity (solid line), neutrino luminosity
(three-dotted-dashed line), nuclear luminosities due to hydrogen burning (dot-dashed line), helium
burning (dotted line), carbon burning (long dashed line)  and DM burning (short dashed line) in the
non-rotating $100~\mathrm{M_\odot}$ model sequence.  
The contribution of DM annihilation to the neutrino luminosity, which
corresponds to $1/3L_\mathrm{DM}$, is not included here.  The rapid increase in $L_\mathrm{H}$ and
$L_\mathrm{He}$ after helium exhaustion is due to the hydrogen and helium shell burning. \emph{Lower
Panel:}  Evolution of the WIMP capture rate ($C_*$, solid line), the thermalization rate
($\Gamma_\mathrm{th}$, dashed line) and the number of thermalized WIMPs ($N_\mathrm{th}$, dotted
line) in the corresponding model. See Eqs.~(2), (3) and (4). 
}\label{fig2} 
\end{figure}

No meaningful change in the stellar structure according to different values of $\rho_\chi$ ($<
\rho_\mathrm{\chi,crit}$) is observed in non-rotating models.  Since DM burning only occurs within a
very small radius $r_\chi (<< R_\mathrm{core})$, stars with different $\rho_\chi$ at a given mass
have similar amounts of energy flux from the core and produce  helium  cores of a similar size (e.g.
$\sim 40~\mathrm{M_\odot}$ from $100~\mathrm{M_\odot}$ stars).  The luminosity resulting from DM
burning gradually increases early on the main sequence as the star expands, but continuously
decreases in later stages since the significant reduction of the number of hydrogen atoms lowers the
DM capture rate.  Rapid increase of the stellar radius after helium exhaustion makes the
thermalization time very long, leading to reduction of the number of thermalized WIMPs (see
Fig.~\ref{fig2}).  The DM luminosity accordingly decreases further from $ 7 \times
10^5~\mathrm{L_\odot}$ to about $10^5~\mathrm{L_\odot}$ during the carbon burning phase,  in the
given example with $100~\mathrm{M_\odot}$ (Fig.~\ref{fig2}).  Carbon burning and particularly
neutrino cooling ($L_\mathrm{\nu} > 10^{10}~\mathrm{L_\odot}$) dominate the evolution at this stage
as shown in Fig~\ref{fig2}.  As the evolution of the star beyond carbon exhaustion should also be
governed by neutrino cooling and other nuclear reactions such as oxygen burning, the effect of DM
burning on the pre-supernova structure must be minor.  As the situation remains similar in the other
models of $20 \le M/\mathrm{M_\odot} \le 300$, we conclude that DM burning may not change the final
fate of the non-rotating first stars. 

\begin{figure}[t] 
\epsscale{1.00} 
\plotone{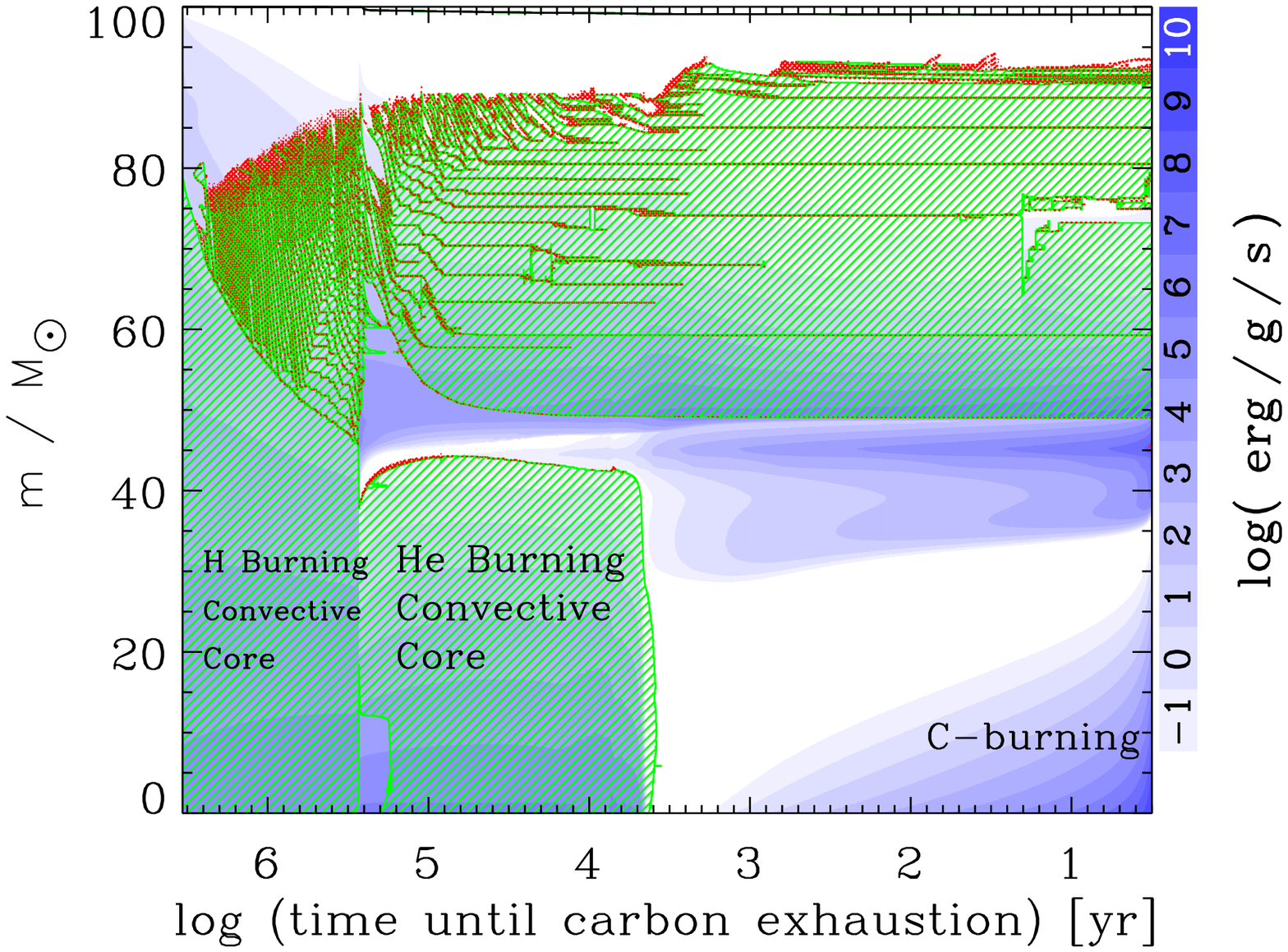} 
\plotone{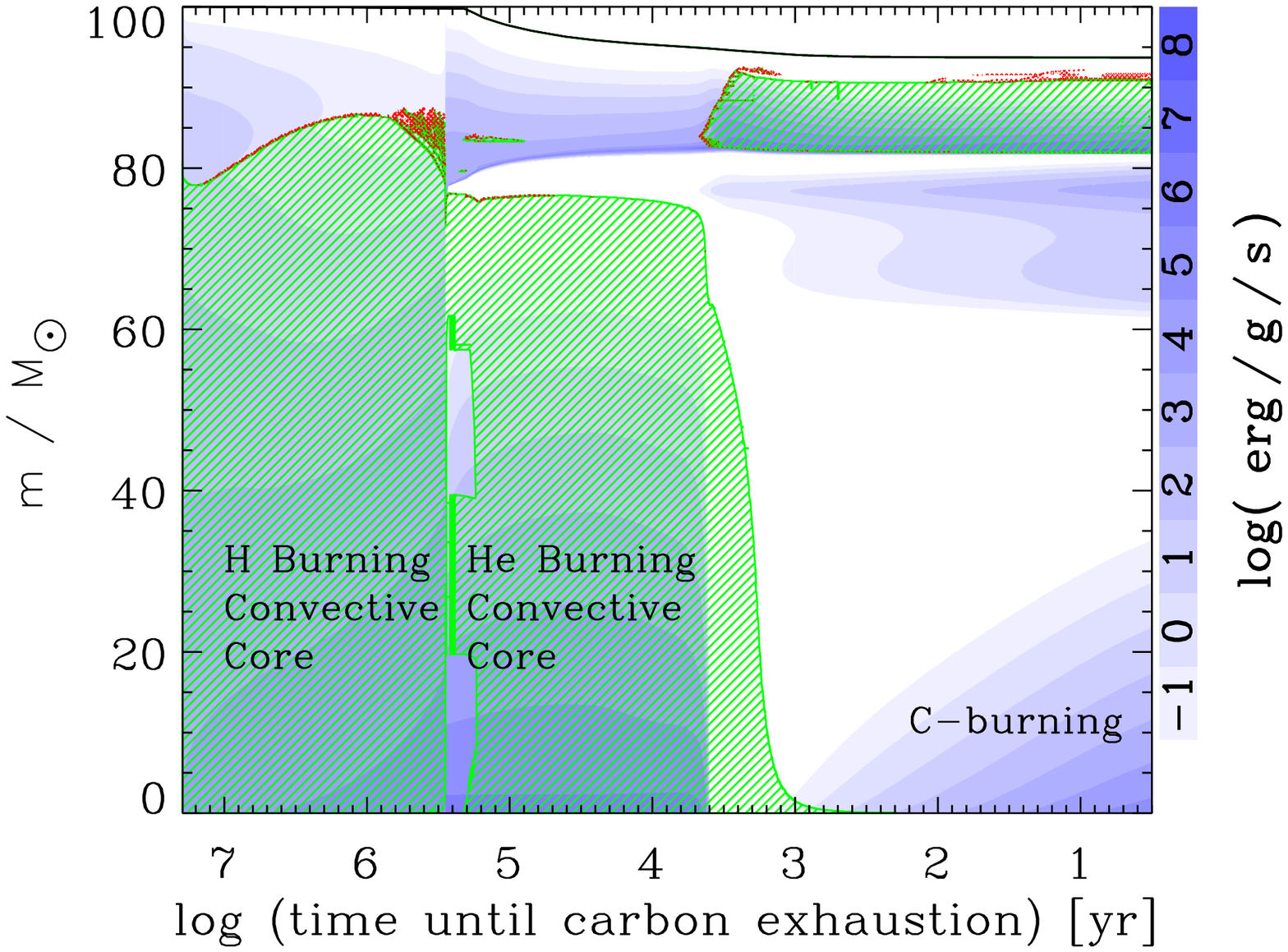} 
\caption{
Evolution of the internal structure of rotating $100~\mathrm{M_\odot}$ star models without DM
burning (upper panel) and with DM burning (lower panel, $\rho_\chi = 4\times10^{10}~\mathrm{GeV
cm^{-3}}$).  Convective layers are hatched, and semi--convective layers are marked by red dots. The
color shading indicates nuclear energy generation rates.  The adopted initial rotational velocity at
the equatorial surface is $132~\mathrm{km~s^{-1}}$, which corresponds to 10 \% of the Keplerian
value.
}\label{fig3} 
\end{figure} 
 
It is noteworthy, however, that rotation can dramatically change  the evolution with DM burning.
Fluids in rotating massive stars are subject to various rotationally induced hydrodynamic
instabilities such as Eddingon-Sweet circulations, which can cause mixing of chemical species across
the boundary between the hydrogen burning core and the radiative envelope.  Such mixing is usually
stabilized by the strong buoyancy potential due to the chemical stratification between the hydrogen
burning core and the envelope.  However, if chemical mixing can occur faster than the building-up of
chemical gradients by nuclear burning (i.e., $\tau_\mathrm{mix}/\tau_\mathrm{nuc} < 1$), then
(quasi-) chemical homogeneity can be maintained on the main sequence (so-called chemically
homogeneous evolution)~\citep{Maeder87}.  The condition, $\tau_\mathrm{mix}/\tau_\mathrm{nuc} < 1$,
can be met either by reducing $\tau_\mathrm{mix}$ or increasing $\tau_\mathrm{nuc}$.  Rapid rotation
tends to do the former, and we find that DM burning tends to do the latter.  This is because the
mixing time scale due to Eddington--Sweet circulations remains almost unchanged with increasing
$\rho_\chi$ (for $ \rho_\chi < \rho_\mathrm{\chi, crit}$), while the nuclear burning time scale
increases due to the DM burning.  This explains the remarkable impact of DM burning in the evolution
of our rotating models as shown in Fig.~\ref{fig3}.  As shown in the upper panel, rotation in this
model ($v_\mathrm{init}/v_\mathrm{Kepler} = 0.1$) is not rapid enough to cause the strong mixing by
itself without DM burning.  With $\rho_\chi = 4 \times 10^{10}~\mathrm{GeV~cm^{-3}}$ (see lower
panel), however, the $100~\mathrm{M_\odot}$ star lives about 10 times longer than in the
corresponding non--DM burning case, and the star undergoes the quasi- chemically homogeneous
evolution even with slow initial rotation because of the DM burning effect on nuclear burning time
scale \citep[cf.][]{Yoon06}.  The star is thus gradually transformed into a massive helium star by
the end of main sequence as almost all of the hydrogen atoms in the star are fused into heliums due
to mixing.

\section{Discussion}\label{sect:discussion} 

Our results indicate that DM burning should not significantly alter our view  on the final fate of
the non--rotating first stars: pair--instability supernovae for $ 140 \la  M/\mathrm{M_\odot} \la
260$, and core--collapse events for other masses \citep{Heger02}, although their life times may be
significantly prolonged.  However, the impact of DM burning appears more important for rotating
stars.  The quasi-chemically homogeneous evolution can be rather easily realized even with moderate
rotation velocities with $10^{10} \la \rho_\chi~[\dmrho] \la 10^{11}$.  Such evolution can lead to
production of massive helium stars that emit large amounts of helium ionizing photons, as well as
abundant production of primary nitrogen, as discussed in \citet{Yoon08}.  Note also that the
quasi-chemically homogeneous evolution scenario (CHES) is one of the favored ones for the production
of long GRBs from metal poor stars~\citep{Yoon05, Woosley06}.  Our result therefore
indicates that DM burning might promote the production of long gamma--ray bursts from the first
stars of $12 \la M/\mathrm{M_\odot} \la 60$ via the CHES channel \citep[see][]{Yoon06}. 
 
DM burning in the first stars must have consequences in the history of reionization in the early
universe.  Table~1 lists the number of hydrogen and helium ionizing photons emitted from
$100~\mathrm{M_\odot}$  models.   If $\rho_\chi \la 2\times10^{11} \dmrho$, the total number of
ionizing photons increases proportionally to the DM density, as a direct consequence of the
life--prolonging effect of DM burning. For a given $\rho_\chi$ rotation does not significantly alter
hydrogen ionizing photon counts nor the lifetime.  However, for models that take path of the CHES,
the helium ionizing photon counts is increased by more than a factor of 2 compared to the non--rotating
case.  If $\rho_\chi$ is very large, on the other hand,  the surface temperature of the star
drops significantly enough (Fig. 1) that the  total number of ionizing photons is reduced even with
much longer  lifetimes due to the DM burning.  For example,  with $\rho_{\chi}= 2\times10^{12}~
\dmrho$, it would have to take $\sim$ 10 Gyr to emit as many hydrogen ionizing photons as a non--DM
burning counterpart.  Therefore, if most of the first stars  had been born with such high
$\rho_\chi$ their contribution to reionization would have been dramatically reduced.  The effect of
the temperature drop on the number of helium ionizing photons is even more prominent: with
$\rho_\chi = 2\times10^{12}~\dmrho$, it decreases by more than 19 orders of magnitude  compared to
the other cases with lower $\rho_\chi$.  Future study on the history of helium ionization at high
redshift might therefore be a strong probe of DM burning in the first stars. 

In this study we assume that the background DM halo density stays constant throughout the stellar
evolution.  This assumption may be valid if the stellar life times are shorter than about 100 Myr --
which is the expected merger timescale of DM halos at  at z $\sim$ 20~\citep[e.g.,][]{Lacey93}  --
as in our model sequences with $\rho_\chi \la 4\times10^{10}~\dmrho$.  The evolution of DM burners
in halos with higher $\rho_\chi$, however, should be critically determined by the change of DM halo
environments.  If $\rho_\chi$ is sufficiently reduced due to merger events and/or to displacement of
the star from the densest region of the DM halo, the DM burners will  become ``normal'' stars,
dominated by nuclear burning.  The star may then die quickly as we expect for normal stars, or
become a ``born--again'' DM burner if $\rho_\chi$ increases again in later stages for some reason.
The detailed history of the feedback from the first stars (e.g. metal enrichment and
reionization) on the evolution of the early universe may depend on the nature and evolution of DM
halos where the first stars are formed.  This issue should be addressed in future work.

\begin{table} 
\begin{center} 
\caption{Number of hydrogen and helium ionizing photons from $100~\mathrm{M_\odot}$ star models}~\label{tab1} 
\begin{tabular}{crccc} 
\tableline \tableline  
$v_\mathrm{rot}/v_\mathrm{K}$ & $\rho_\chi~[\dmrho]$  &  $N_\mathrm{H}$  & $N_\mathrm{He}$  & Duration \\ 
\tableline 
    0.0               &      0.00     &  $1.2 \times 10^{64}$  & $2.2 \times 10^{62}$    & 3.2 Myr \\ 
    0.0               &      $2\times10^{10}$     &  $2.1 \times 10^{64}$  & $3.4 \times 10^{62}$    & 5.5 Myr \\ 
    0.0               &      $4\times10^{10}$     &  $8.5 \times 10^{64}$  & $1.5 \times 10^{63}$    & 22.3 Myr \\ 
    0.0               &      $10^{11}$     &  $2.9 \times 10^{65}$  & $6.3 \times 10^{62}$    & 100.0 Myr\tablenotemark{*} \\ 
    0.0               &      $2\times10^{11}$     &  $2.0 \times 10^{65}$  & $1.7 \times 10^{61}$    & 100.0 Myr\tablenotemark{*} \\ 
    0.0               &      $2\times10^{12}$     &  $1.3 \times 10^{62}$  & $9.4 \times 10^{43}$    & 100.0 Myr\tablenotemark{*} \\ 
\tableline
    0.1               &      0.00    &  $1.5 \times 10^{64}$  & $2.5 \times 10^{62}$    & 3.4 Myr \\ 
    0.1               &      $2\times10^{10}$ &  $2.7 \times 10^{64}$  & $5.2 \times 10^{62}$    & 6.0 Myr \\ 
    0.1               &      $4\times10^{10}$ &  $8.7 \times 10^{64}$  & $3.8 \times 10^{63}$    & 19.6 Myr \\ 
\tableline 
\end{tabular} 
\vspace{-0.5mm}
\tablenotetext{*}{The numbers are calculated  only for the first 100 Myr.} 
\end{center} 
\end{table}

\acknowledgements
As a note added in proof, we acknowledge that \citet{Taoso08} independently
report similar results about the effect of DM burning on the MS life time and
some stellar properties, using a different numerical code.  S.~A. appreciates
helpful discussion with M. Alvarez and M.Busha.  S.~A. is supported by a
Department of Energy Contract to SLAC DE--AC3--76SF00515.  F.~I. is grateful to
P.  Scott and M. Taoso for helpful discussion.  F.~I. is supported by MIUR
through grant PRIN--2006.  S.C.~Y. is grateful to C. Church for the help with
the text, and to W. Hillebrandt and S. Woosley for supporting his visit to MPA,
Garching, in June, 2008, where part of the manuscript has been prepared.
S.C.~Y. is supported by the DOE SciDAC Program (DOE DE-FC02-06ER41438).

\end{document}